\documentstyle[graphicx,aps,multicol]{revtex}
\draft

\begin{document}

\title{Evolution of level density step structures from $^{56,57}$Fe to 
$^{96,97}$Mo}
\author{A.~Schiller,$^{1}$\footnote{Electronic address: 
Andreas.Schiller@llnl.gov}, E.~Tavukcu,$^{1,2,3}$\footnote{Present address: 
Osmangazi University, Dept.\ of Physics, Meselik, Eskisehir, 26480 Turkey} 
L.A.~Bernstein,$^1$ P.E.~Garrett,$^1$ M.~Guttormsen,$^4$ M.~Hjorth-Jensen,$^4$ 
C.W.~Johnson,$^5$ G.E.~Mitchell,$^{2,3}$ J.~Rekstad,$^4$ S.~Siem,$^4$ 
A.~Voinov,$^6$ W.~Younes$^1$}
\address{$^1$ Lawrence Livermore National Laboratory, L-414, 7000 East Avenue, 
Livermore, California 94551, USA}
\address{$^2$ North Carolina State University, Raleigh, North Carolina 27695, 
USA}
\address{$^3$ Triangle Universities Nuclear Laboratory, Durham, North Carolina 
27708, USA}
\address{$^4$ Department of Physics, University of Oslo, N-0316 Oslo, Norway}
\address{$^5$ San Diego State University, San Diego, California 92182, USA}
\address{$^6$ Frank Laboratory of Neutron Physics, Joint Institute of Nuclear 
Research, 141980 Dubna, Moscow region, Russia}

\maketitle

\begin{abstract}
Level densities have been extracted from primary $\gamma$ spectra for 
$^{56,57}$Fe and $^{96,97}$Mo nuclei using ($^3$He,$\alpha\gamma$) and
($^3$He,$^3$He$^\prime\gamma$) reactions on $^{57}$Fe and $^{97}$Mo targets. 
The level density curves reveal step structures above the pairing gap due to 
the breaking of nucleon Cooper pairs. The location of the step structures in 
energy and their shapes arise from the interplay between single-particle 
energies and seniority-conserving and seniority-non-conserving interactions.
\end{abstract}


\begin{multicols}{2}

The group at the Oslo Cyclotron Laboratory (OCL) has recently developed a new 
method (the so-called Oslo method) to extract the level density and radiative 
strength function from primary $\gamma$ spectra \cite{SB00}. The method can be
characterized as a further development of the sequential extraction method 
\cite{BB70,BE73}. The Oslo method has been extensively tested in the rare-earth
mass region \cite{SB01,MG01,VG01,SG02} and has been successfully extended to 
the very light $^{27,28}$Si nuclei \cite{GM02}. The robustness of the Oslo 
method in the mid \it sd \rm shell where the level density is rather low and 
the single-particle levels are widely spaced prompted the present 
investigations in mass regions around the $f_{7/2}$ shell (the iron nuclei) and
around the \it fpg \rm shell (the molybdenum nuclei). The $^{56}$Fe nucleus was
especially chosen due to astrophysical interest in this isotope. While Ref.\
\cite{BB57} discusses the $e$-process and the direct production of $^{56}$Fe, a
modern view in terms of a hierarchy of restricted nuclear statistical 
equilibria (NSE) on different time scales predicts the direct production of 
$^{56}$Fe only on the highest level of NSE, namely in the presence of weak 
equilibrium \cite{WI97}. These rather extreme conditions are envisioned only in
the core of massive stars (where ejection into space is unlikely) and in the 
core of Type Ia supernovae (where ejection into space seems possible). In the 
equations of NSE, the nuclear partition function, i.e., the Laplace transform 
of the nuclear level density, plays a prominent role. The nucleus $^{96}$Mo is 
of special interest in the investigation of the $|N-Z|$ dependence of level 
densities, since the mass 96 isobars are the only ones in the nuclear chart 
where one can find three different stable nuclei with $|N-Z|$ varying by eight 
units from $^{96}$Zr to $^{96}$Ru. Also the $|N-Z|$ dependence of level 
densities can have significant astrophysical importance \cite{AG01}, since many
reactions take place on unstable isotopes and therefore most of their reaction 
cross sections must be estimated by Hauser-Feshbach type calculations 
\cite{HF52}.

The purpose of this Rapid Communication is to report on experimental level
densities of $^{56,57}$Fe and $^{96,97}$Mo nuclei and to provide a schematic 
explanation of the observed step structures in these curves.

\begin{figure}\centering
\includegraphics[totalheight=17.9cm]{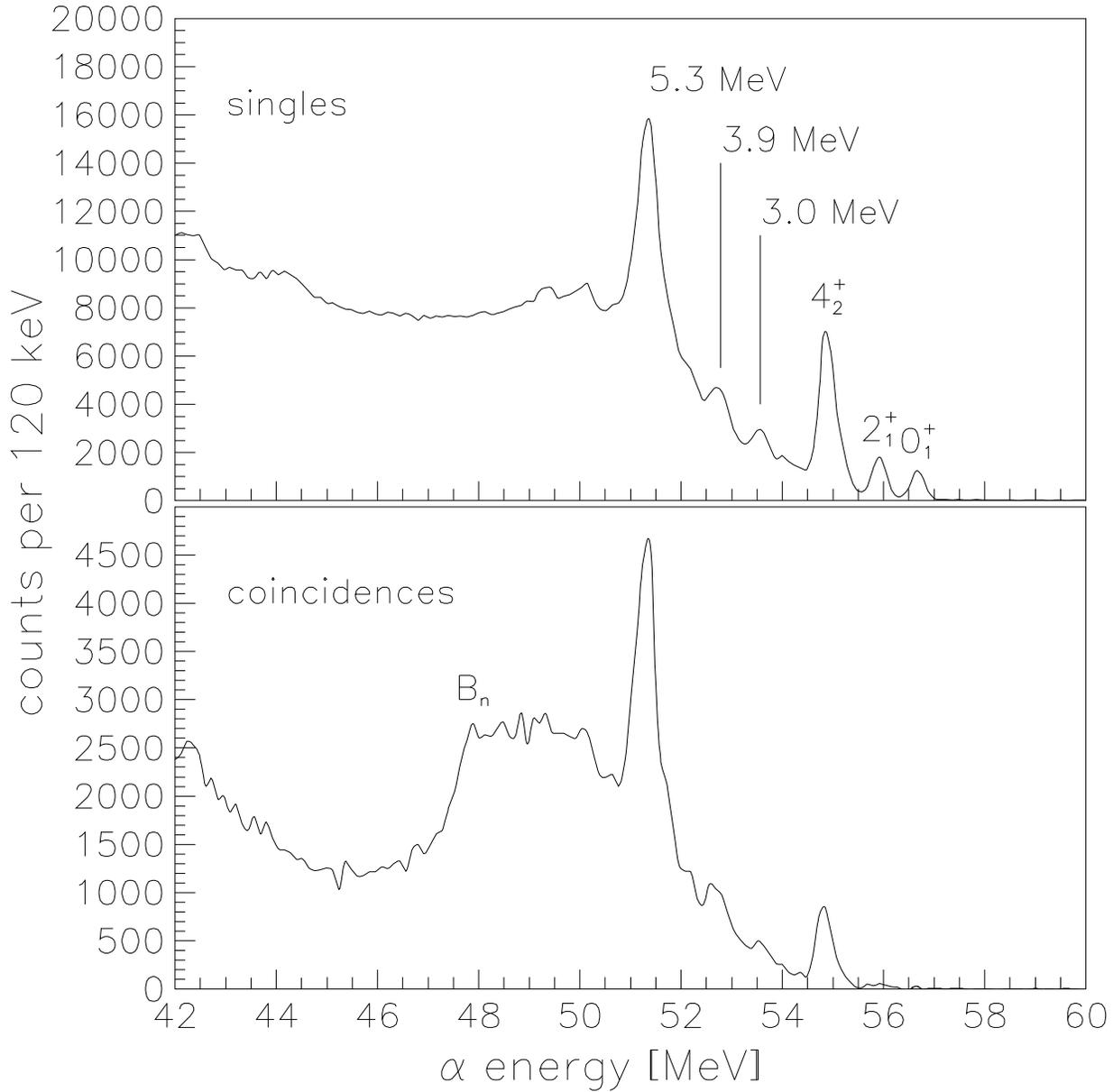}
\caption{A singles $\alpha$ spectra is shown in the upper panel and an 
$\alpha$-$\gamma$ coincidence spectrum is shown in the lower panel for the 
$^{97}$Mo($^3$He,$\alpha$)$^{96}$Mo reaction. At the neutron binding energy 
$B_n$, the coincidence spectrum shows a dip reflecting the lower $\gamma$ 
multiplicity in the decay from low-lying states in $^{95}$Mo which are 
populated by neutron emission. The transfer peaks to the first three states are
well known from the $(p,d)$ reaction \protect\cite{CM73}. The strong transfer 
peak at 5.3~MeV is a new discovery and indicates the opening of the neutron 
$g{9/2}$ shell at this energy.}
\label{fig:moraw}
\end{figure}

The experiments were performed at the MC-35 cyclotron at the OCL using a 
$\sim 2$~nA beam of 45~MeV $^3$He particles. The self-supporting targets were 
isotopically enriched to 94.7\% and 94.2\% and had thicknesses of 3.4~mg/cm$^2$
and 2.1~mg/cm$^2$ for $^{57}$Fe and $^{97}$Mo, respectively. Each experiment 
ran for approximately five days each, and about 200,000 relevant 
particle-$\gamma$ coincidences were recorded in each analyzed reaction channel.
The charged ejectiles were identified and their energies measured in a ring of 
eight collimated Si $\Delta E$-$E$ telescopes placed at 45$^\circ$ with respect
to the beam direction. The thicknesses of the front and end detectors were 140 
and 3000~$\mu$m, respectively and shielding against $\delta$ electrons was 
achieved with a 19-$\mu$m-thick Al foil. The distance from the target was 5~cm,
giving a total solid angle coverage of 0.3\% of $4\pi$, and the energy 
resolution was $\sim$0.3~MeV over the entire spectrum. The $\gamma$-rays were 
detected in 28 collimated 5"x5" NaI(Tl) detectors called CACTUS \cite{GA90} 
surrounding the target and particle detectors. The total efficiency of CACTUS 
was $\sim$15\% of $4\pi$ and the resolution was $\sim$6\% of the deposited 
energy at 1.3~MeV\@. In addition, one 60\% Ge(HP) detector was used in the 
setup to monitor the selectivity and populated spin distribution of the 
reactions. Raw $\alpha$-particle data are shown in Fig.\ \ref{fig:moraw} for 
the case of the $^{97}$Mo($^3$He,$\alpha$)$^{96}$Mo reaction, where discrete
transfer peaks are observed up to $\sim$6~MeV in excitation energy indicating 
the opening of the neutron $g_{9/2}$ shell.

\begin{figure}\centering
\includegraphics[totalheight=11.9cm]{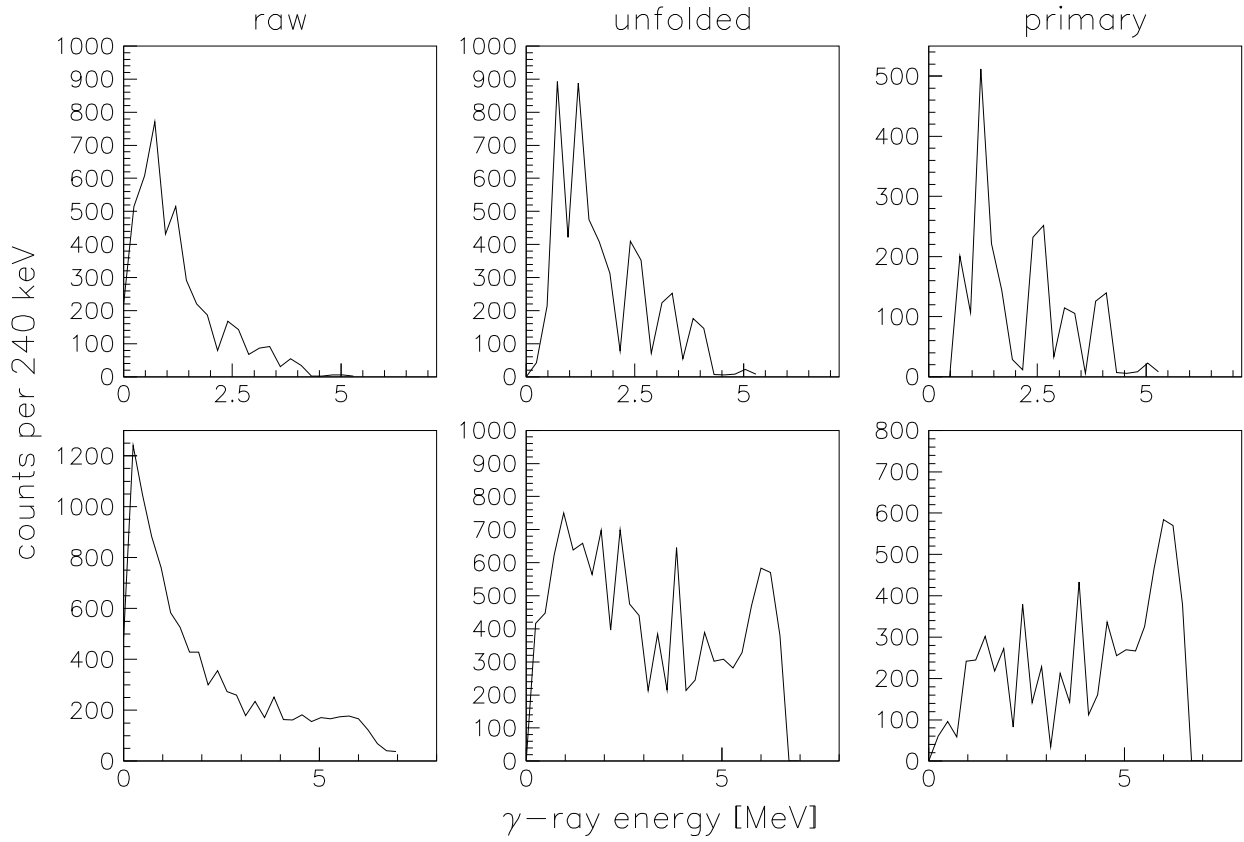}
\caption{Raw, unfolded, and primary $\gamma$ spectra from the 
$^{57}$Fe($^3$He,$\alpha\gamma$)$^{56}$Fe reaction at 5~MeV excitation energy 
(upper panels) and from the  $^{57}$Fe($^3$He,$^3$He$^\prime\gamma$)$^{57}$Fe 
reaction at 6.2~MeV excitation energy (lower panels).}
\label{fig:feraw}
\end{figure}

From the known $Q$-value and the reaction kinematics, the ejectile energy can 
be transformed into initial excitation energy of the residual nuclei. Using the
particle-$\gamma$ coincidence technique, each $\gamma$ ray can be assigned to
a cascade depopulating a certain initial excitation energy in the residual 
nucleus. The data are therefore sorted into total $\gamma$-ray spectra 
originating from different initial excitation energy bins. Every spectrum is
then unfolded using a Compton-subtraction method which preserves the 
fluctuations in the original spectra and does not introduce further, spurious 
fluctuations \cite{GT96}. From the unfolded spectra, a primary $\gamma$ matrix
is constructed using the subtraction method of Ref.\ \cite{GR87}. The basic 
assumption behind this method is that the $\gamma$-ray decay pattern from any 
excitation energy bin is independent of whether states in this bin are 
populated directly via the ($^3$He,$\alpha$) or ($^3$He,$^3$He$^\prime$) 
reactions or indirectly via $\gamma$ decay from higher excited levels following
the initial nuclear reaction. This assumption is trivially fulfilled if one
populates the same levels with the same weights within any excitation energy 
bin, since the decay branchings are properties of the levels and do not depend 
on the population mechanisms. On the other hand, if one populates, e.g., vastly
different spin distributions within one excitation energy bin via the two 
different population mechanisms, the $\gamma$-decay patterns should be 
different as well. As an example of the data analysis discussed in this 
paragraph, the raw, unfolded and primary $\gamma$ spectra are shown for the 
$^{57}$Fe($^3$He,$\alpha\gamma$)$^{56}$Fe and 
$^{57}$Fe($^3$He,$^3$He$^\prime\gamma$)$^{57}$Fe reactions in Fig.\ 
\ref{fig:feraw}.

\begin{figure}\centering
\includegraphics[totalheight=8.9cm]{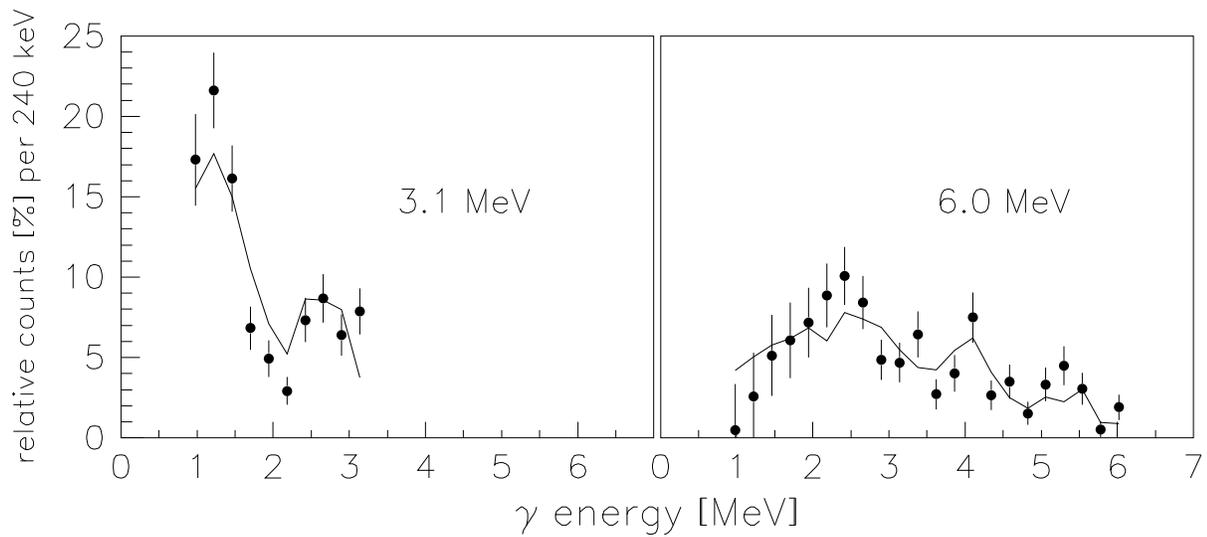}
\caption{Experimental primary $\gamma$ spectra (data points with error bars) at
two different initial excitation energies (indicated in the figure) compared to
the least $\chi^2$ fit (solid lines) for the 
$^{97}$Mo($^3$He,$^3$He$^\prime\gamma$)$^{97}$Mo reaction. The fit is performed
simultaneously to the whole primary $\gamma$ matrix of which the two displayed 
spectra are only a small fraction. The first generation spectra are normalized
to one at each excitation energy.}
\label{fig:fgmo}
\end{figure}

Finally, the primary $\gamma$ matrix is factorized using the generalized 
Brink-Axel hypothesis \cite{Br55,Ax62}. The original hypothesis states that the
giant dipole resonance (GDR) can be built on every excited state, and that the 
properties of the GDR do not depend on the temperature of the nuclear state on 
which it is built. This hypothesis can be generalized to include not only the 
GDR but any type of nuclear excitation and results in the assumption that 
primary $\gamma$ spectra originating from the excitation energy $E$ can be 
factorized into a $\gamma$-ray transmission coefficient 
${\mathcal{T}}(E_\gamma)$ which depends only on the $\gamma$-transition energy 
$E_\gamma$ and into the level density $\rho(E-E_\gamma)$ at the final energy. 
This factorization is determined by a least $\chi^2$ fit to the primary 
$\gamma$ matrix, using no \sl a priori \rm assumptions about the functional 
form of either the level density or the $\gamma$-ray transmission coefficient 
\cite{SB00}. An example to illustrate the quality of the fit is shown in Fig.\ 
\ref{fig:fgmo}, where for the $^{97}$Mo($^3$He,$^3$He$^\prime\gamma$)$^{97}$Mo 
reaction we compare the experimental primary $\gamma$ spectra from two 
different initial excitation energies to the least $\chi^2$ fit. Unfortunately,
the mathematical structure of the relevant equations in the least $\chi^2$ fit 
does not allow us to find a unique solution for the level density and 
$\gamma$-ray transmission coefficient. However, it has been shown that all 
solutions with the same $\chi^2$ can be obtained by the transformation of one 
randomly chosen solution according to \cite{SB00}
\begin{eqnarray}
\tilde{\rho}(E-E_\gamma)&=&A\exp[\alpha(E-E_\gamma)]\rho(E-E_\gamma)\\
\tilde{\mathcal{T}}(E_\gamma)&=&B\exp(\alpha E_\gamma){\mathcal{T}}(E_\gamma).
\end{eqnarray}
The three free parameters $A$, $B$, and $\alpha$ have to be determined to give 
the physically most relevant solution to the least $\chi^2$ fit using 
independent experimental information. The most common way is to count the 
number of discrete levels at low excitation energies and to use the neutron 
resonance spacing at $B_n$ to find values for $A$ and $\alpha$. The remaining
parameter $B$ is then determined using the average total radiative width of 
neutron resonances \cite{VG01}. Unfortunately, in the case of $^{56}$Fe, there
are no data on neutron resonances and thus, the information about the level 
density around $B_n$ in $^{56}$Fe has to be obtained by different means. In
order to do so, we calculate the level density at $B_n$ in $^{57}$Fe using a
backshifted Fermi-gas expression with the parameterization of von Egidy \sl et 
al.\ \rm \cite{ES88}, where we apply an additional overall re-normalization 
factor to match the level density determined from neutron resonance spacings. 
Then, we use the same level-density formula (including the same 
re-normalization factor but with the other parameters for $^{56}$Fe) to 
calculate the level density at $B_n$ in $^{56}$Fe. Using this data point 
instead of the unknown neutron resonance spacing, we proceed in the same way as
for the other three nuclei \cite{SB00,Ta02}. 

\begin{figure}\centering
\includegraphics[totalheight=8.9cm]{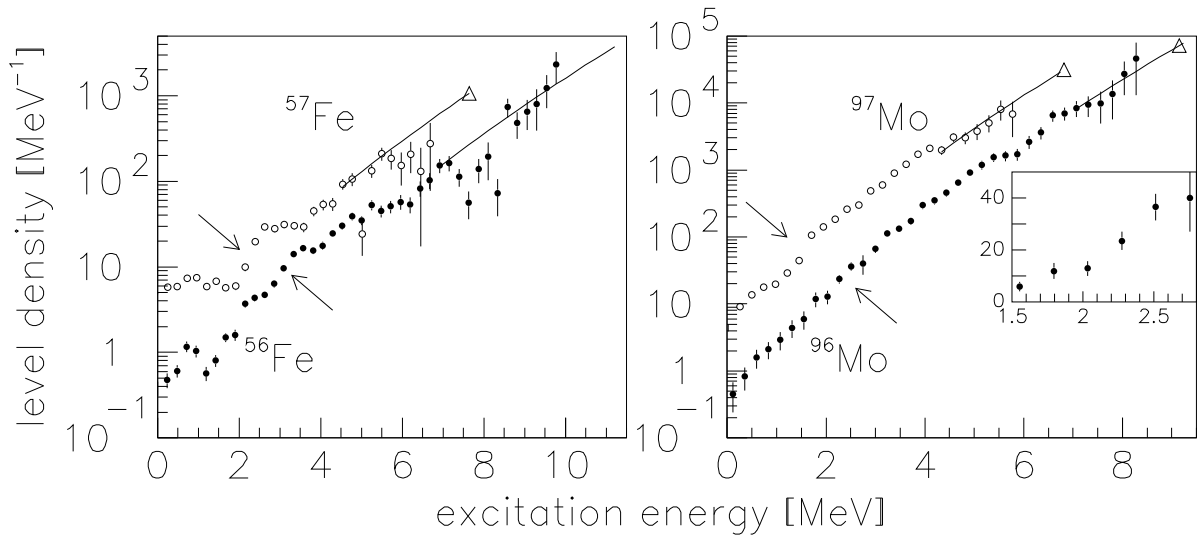}
\caption{Experimental level densities for the four nuclei under study. Where
available, level density data from neutron resonance spacings have been added 
(open triangles). The solid lines are the re-normalized level density 
parameterizations according to von Egidy \sl et al.\ \rm \protect\cite{ES88}. 
Step structures in the level densities are marked by arrows. The insert shows 
the step structure for $^{96}$Mo on a linear scale. In the level density of 
$^{56}$Fe, the bump and the plateau at 0.8~MeV and 2.0~MeV, respectively, are 
due to the first and second excited states.}
\label{fig:levdens}
\end{figure}

In Fig.\ \ref{fig:levdens} we show the physically most relevant solutions for 
the level densities in $^{56,57}$Fe and $^{96,97}$Mo from the least $\chi^2$ 
fit. The most striking feature in those curves are the steps starting at 
2.9~MeV and 1.8~MeV in $^{56}$Fe and $^{57}$Fe. There are also less pronounced 
but still statistically significant step structures at 2.0~MeV and 1.2~MeV in 
$^{96}$Mo and $^{97}$Mo. It has been established in the rare-earth region that 
such low-lying step structures are connected to the breaking of the first 
nucleon Cooper pair \cite{MB99}. It is therefore natural to assume that the 
same is true for the lighter mass regions. The additional step at 6~MeV in the 
level density curve of $^{96}$Mo might be due to the opening of the $g_{9/2}$ 
shell which is indicated to happen at this excitation energy (see Fig.\ 
\ref{fig:moraw}). A future systematic investigation of Mo nuclei might shed 
more light on this issue.

The difference in binding energy between the even and odd systems is a measure 
for pairing correlation and can be calculated from the three-mass indicator of 
Dobaczewski \sl et al.\ \rm \cite{DM01}. This indicator yields 1.2~MeV and 
0.9~MeV for the Fe and Mo nuclei, respectively, which agrees very well with the
differences in excitation energy for the first steps in the level density 
curves. Thus we know that the steps in neighboring nuclei appear at the same 
'effective' excitation energy, i.e., the excitation energy corrected for the 
contribution to the binding energy due to neutron-pairing correlations. 
Further, the average proton pairing energies are 0.8~MeV and 1.1~MeV for the 
$^{57}$Fe and $^{97}$Mo nuclei, respectively. These energies should now 
correspond to the excitation energies of the steps in the two odd nuclei. 
However, the steps are delayed in excitation energy by 1.0~MeV and 0.1~MeV for 
these two nuclei. This might be explained by the fact that one not only has to 
invest the energy to break a proton Cooper pair but also at least one of the 
unpaired protons has to be promoted to the next unoccupied single-particle 
level. The spacing of those levels can also be calculated using the Dobaczewski
three-mass indicator, giving 2.0~MeV and 0.9~MeV in the two cases. Clearly, the
higher single particle spacing for the $^{56}$Fe nucleus leads to a larger 
delay in excitation energy for the step structure to appear compared to the 
$^{97}$Mo nucleus, but obviously the exact excitation energy for the steps in 
the level density curves cannot be estimated from binding energies, since it 
will depend on the exact location of the Fermi energy within the 
single-particle level scheme.

\begin{figure}\centering
\includegraphics[totalheight=8.9cm]{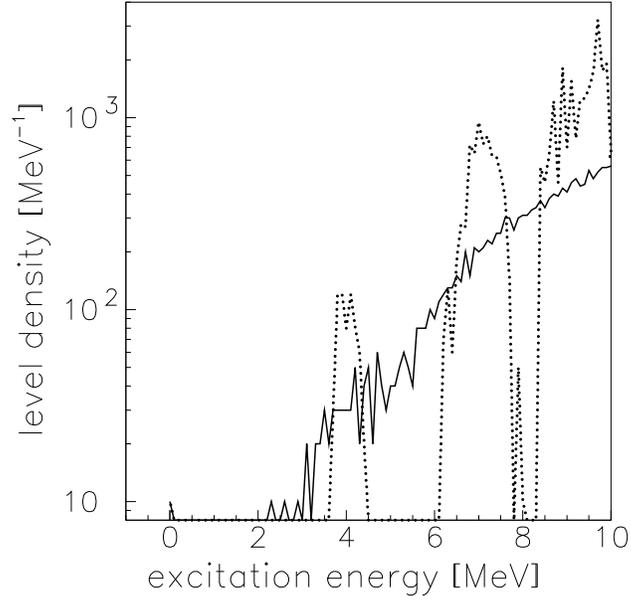}
\caption{Model calculation using the Hamiltonian of Eq.\ (\protect\ref{eq:ham})
with $\epsilon=0.25$~MeV, $G=0.5$~MeV, and $\kappa=0$~MeV (dotted line). The 
level density is given as function of excitation energy. Adding a random 
two-body interaction with the strength $\kappa=0.14$~MeV (solid line) results 
in a step structure quite similar to what is seen in the experimental level 
density curve of the iron nuclei.}
\label{fig:calc}
\end{figure}

Another complication might arise from the effect of seniority non-conserving
interactions, which mix configurations of different seniority and smooth out 
the step structures in the level densities. To investigate this, we have 
performed a model calculation. In the model we assume a system of eight 
particles scattered into an equidistant single-particle level scheme with eight
doubly-degenerate levels. As residual interactions we consider a pairing 
interaction and a seniority non-conserving interaction. Thus, the model 
Hamiltonian is written as
\begin{eqnarray}
\lefteqn{\widehat{H}=\epsilon\sum_{i=1}^8ia_i^\dagger a_i-\frac{1}{2}G
\sum_{i,j=1}^8a_i^\dagger a_{\bar{\imath}}^\dagger a_{\bar{\jmath}}a_j}
\nonumber\\
&&-\frac{1}{2}\kappa\sum_{i,j,k,l=1}^8W_{ijkl}a_i^\dagger a_j^\dagger 
a_ka_l,
\label{eq:ham}
\end{eqnarray}
where $a^\dagger$ and $a$ are Fermion creation and annihilation operators and 
the labels with bars stand for time reversed orbits. The single particle level
spacing $\epsilon$, the strength $G$ of the pairing interaction and the 
strength $\kappa$ of the seniority non-conserving interaction $W$ are the only 
macroscopic parameters of the model. This model with good seniority, i.e., the 
case of $\kappa=0$ has already been diagonalized in Ref.\ \cite{GB00}, and the 
dotted line in Fig.\ \ref{fig:calc} gives the distribution of eigenvalues with 
excitation energy, i.e., the exact level density of the model. The individual 
bumps contain mainly levels with the same seniority, thus the step structures 
in the level density can be explained by the consecutive breaking of nucleon 
Cooper pairs. However, the experimental data show that in general the steps are
much smoother than in such a simple model calculation indicating the need for 
seniority non-conserving terms in the Hamiltonian.

An important contribution to seniority breaking comes from quadrupole 
collectivity, which can change rapidly with mass number for the nuclei under 
study. Strong, residual quadrupole-quadrupole interactions can lead to 
deformation of the nucleus and to a modification of the single-particle level 
spacing. Unfortunately, our simple model cannot incorporate a realistic 
quadrupole-quadrupole interaction; the smoothing of the level density, however,
should not depend on the details of the seniority non-conserving residual 
interaction. We choose therefore to model the quadrupole-quadrupole interaction
by a random two-body interaction \cite{MF75} of roughly equivalent strength 
where all the pairing-like terms have been set to zero i.e., the $W_{ijkl}$ in
Eq.\ (\ref{eq:ham}) are Gaussian random numbers of mean zero and width equal to
one except for the cases of $j=\bar{\imath}$ and $k=\bar{l}$ where
$W_{i\bar{\imath}\bar{l}l}=0$.

Since in this more general case Eq.\ (\ref{eq:ham}) does not have good 
seniority, exact diagonalization of the model Hamiltonian has to be performed 
within the full model space which is computationally very demanding, as one 
requires \it all \rm levels, not just the lowest. For eight particles in eight 
doubly-degenerate states there are 12870 states (counting all magnetic 
substates). The number of levels (not counting magnetic substates) equals the 
number of states with spin projection $J_z=0$, thus yielding 4900 levels. For 
the present calculations, we used $\epsilon=0.25$~MeV, $G=0.5$~MeV, and 
$\kappa\approx 0$~MeV (pure pairing) and $\kappa=0.14$~MeV (pairing+random 
interaction). In Fig.\ \ref{fig:calc}, we show the resulting level densities of
the two calculations. The resolved bumps with definite seniority in the pure 
pairing case are smeared out by the random interaction. The gaps between the 
bumps are rapidly being filled, while the bumps themselves are degraded into a 
step structure quite similar to the step structure seen in the experimental 
level density curves of the iron nuclei. The occurrence of the step structure 
in the calculation is actually quite sensitive to the exact choice of the 
strength of the random interaction. A weak random interaction 
($\kappa\le 0.1$~MeV) does not fill the gaps between the bumps, a strong random
interaction ($\kappa\ge 0.2$~MeV) produces a more smeared out step structure in
the level density curve similar to the Mo data, indicating the presence of 
relatively stronger seniority non-conserving interactions in the Mo nuclei. The
range of values which produces good qualitative agreement with the Fe data is 
about 0.13~MeV$\le\kappa\le 0.16$~MeV\@. However, due to the simplicity of the 
model, no attempt has been made to achieve quantitative agreement between the 
model and the experimental data.

In conclusion, we have presented new experimental data on level densities below
$B_n$ in $^{56,57}$Fe and $^{96,97}$Mo. Step structures in the level density 
curves have been related to the breaking of nucleon Cooper pairs. The 
occurrence of the step structures in energy depends on the exact location of 
the Fermi energy in the single particle level scheme. The smoothness of the 
step structures depends on the strength of the seniority non-conserving 
interactions, e.g., the quadrupole-quadrupole interaction compared to the 
pairing interaction and the average single-particle level spacing. The more 
pronounced step structures in the level density data of Fe nuclei compared to 
Mo indicate relatively stronger seniority non-conserving interactions in the 
latter case. Our schematic calculations where random two-body interactions were
used to model residual seniority non-conserving interactions show that level 
density curves are very sensitive to the relative strengths of the different 
terms in the model Hamiltonian. The present experimental data therefore break
new ground for the possible theoretical investigation of two of the most 
important residual interactions in atomic nuclei, namely the pairing and the 
quadrupole-quadrupole interaction

Part of this work was performed under the auspices of the U.S. Department of 
Energy by the University of California, Lawrence Livermore National Laboratory 
under Contract No.\ W-7405-ENG-48. Financial support from the Norwegian 
Research Council (NFR) is gratefully acknowledged. G.M. acknowledges support
by a U.S. Department of Energy grant with No.\ DE-FG02-97-ER41042.

\end{multicols}

\clearpage

\clearpage

\clearpage

\clearpage

\clearpage

\end{document}